\newcommand \be{\begin{equation}}
\newcommand \ba{\begin{eqnarray}}
\newcommand \ea{\end{eqnarray}}
\newcommand \ee{\end{equation}}

\documentclass[aps,twocolumn,showpacs,preprintnumbers,amsmath,amssymb,floatfix,showkeys]{revtex4}
\usepackage{epsfig}
\usepackage{color}
\usepackage{array}
\begin{document}

\title{Correlation effects in sequential energy branching: an exact model of the Fano statistics}

\author{Arsen V. Subashiev}
\email[]{subashiev@ece.sunysb.edu}
 \author{ Serge Luryi}
 \email[]{Serge.Luryi@stonybrook.edu}
\affiliation{Department of Electrical and Computer Engineering,
State University of New York at Stony Brook, Stony Brook, NY,
11794-2350 }

\begin{abstract}  Correlation effects in the fluctuation of the
number of particles in the process of energy branching by sequential
impact ionizations are studied using an exactly soluble model of
random parking on a line. The Fano factor $F$ calculated in an
uncorrelated final-state ``shot-glass'' model does not give an
accurate answer even with the exact gap-distribution statistics.
Allowing for the nearest-neighbor correlation effects gives a
correction to $F$ that brings $F$ very close to its exact value. We
discuss the implications of our results for energy resolution of
semiconductor gamma detectors, where the value of $F$ is of the
essence. We argue that $F$ is controlled by correlations in the
cascade energy branching process and hence the widely used
final-state model estimates are not reliable --- especially in the
practically relevant cases when the energy branching is terminated
by competition between impact ionization and phonon emission.

\end{abstract}


\maketitle

 \section{Introduction}

Energy resolution of semiconductor gamma detectors relies on the
ability to accurately estimate the energy deposited by the gamma
photon. The measured quantity is the number of electron-hole ({\em
e-h}) pairs produced in subsequent ionization processes. This number
is estimated either from the charge of electrons and holes separated
by the external electric field in diode detectors, or from the
number of lower-energy photons generated in recombination of the
{\em e-h} pairs in scintillators.

The number $N$ of {\em e-h} pairs generated by a gamma particle is
proportional to its energy, $N =E/ \epsilon$, where $\epsilon$ is
the average pair excitation energy. The impact ionization cascade
leading to multiplication of the pair number is referred to as the
sequential energy branching (SEB).

Gamma-ray spectroscopy requires an accurate measurement of $N$. The
spread in this measurement is the ultimate origin of the imperfect
detector energy resolution. If the efficiency of the detector is
very low, $Y \ll 1$ (i.e. when most of the deposited energy is lost
before the creation of all {\em e-h} pairs), the number of created
pairs is a random variable that can be regarded as a sum of
independent contributions corresponding to the small-probability
events of pair production. The Poisson statistics should apply in
this case, so that the average number of pairs $\langle N \rangle$
and the variance of the pair number are related by $\langle \delta
N^2 \rangle=\langle N \rangle$. In the opposite limit of very high
efficiency, $Y \approx 1$, the number of created pairs will not
fluctuate being strictly fixed by the energy conservation law,
$E=\epsilon N $. In this case, the residual loss is essentially
constant for all events.

For experimentally relevant efficiencies, the ratio of pair-number
variance to that expected for Poisson's statistics is called the
Fano factor, \be F = \frac {\langle \delta N^2 \rangle} { \langle N
\rangle}~. \ee Experimentally, $F$ can be substantially less than
unity. The suppression of fluctuations in the number of ionization
processes was first noticed by Ugo Fano in 1947 \cite{Fano}. He
pointed out that the main source of the suppression is correlation
in the energy distribution between the resulting particles due to
the fixed initial energy.

There have been many attempts  to evaluate the Fano factor
theoretically. The most popular approaches are based on simplified
models that estimate the energy spread in the final energy
distribution of secondary {\em e-h} pairs
\cite{DrumMoll,Bilder,Spieler,Klein}. We shall generally refer to
these approaches as the ``final-state models''. These models assume
that ($i$) the energies of secondary particles are statistically
independent variables described by a single-particle distribution
function; and ($ii$) this distribution function is determined by a
microscopic model of energy sampling (e.g., the impact ionization
model specified by the density of states and the scattering matrix
elements) ---  so that it can be calculated independently
\cite{DrumMoll} or even postulated for a particular branching
mechanism \cite{Bilder,Spieler,Klein}.

The values of $F$ calculated in final-state models are often quite
close to the experimentally observed values. However, since this
calculations were based on widely different underlying physical
models, one would be justified to suspect the agreement to be
somewhat fortuitous. Thus, for the case of Ge, similar results were
obtained by either assuming the dominant role of phonon losses
\cite{Spieler} or by neglecting these losses altogether
\cite{Bilder}. In fact, the more accurate attempts to fit experiment
often involve either unrealistic hypotheses on the phonon losses or
the necessity to adjust upward the bandgap of the material (which is
one of the better known values experimentally and should not be used
as an adjustable parameter). Detailed discussion of the published
results can be found in a recent review \cite{Devhan}.

Recently, an attempt was made to justify the final-state model
approach theoretically. Assuming uncorrelated energy sampling in
every impact ionization event and using the central limit theorem,
the authors of \cite{Jordan} arrived at a formula for $F$ that
requires for its evaluation the knowledge of only the one-particle
distribution function. Although evaluation of this function was
beyond the scope of \cite{Jordan}, one could presume that the use of
the true (exact) one-particle distribution function would give an
accurate value of $F$.

This paper inquires into the validity of this presumption on the
basis of an exactly soluble model. It had been pointed out earlier
\cite{Rusb,Alig} that the energy distribution of the secondary
particles produced by energy branching may be highly correlated by
the very nature of branching itself. However, the relative
importance of these correlations in the estimation of $F$ has not
been clarified. As a result, their physics has remained rather
obscure.

Here we examine the correlation effects in the fluctuation of the
number of particles produced by an impact ionization cascade for an
exactly soluble energy branching model, called the random parking
problem (RPP). As discussed earlier \cite{ASSL}, the RPP on a line
is an accurate model of the energy branching by impact ionization in
a semiconductor with narrow valence band and constant
conduction-band density of states. In such a hypothetical
semiconductor, the impact ionization process produces holes with
vanishing kinetic energy and hence the initial energy is shared
between two secondary electrons only. This is exactly similar to the
way parking of a car in the RPP divides the initial gap into two
parts. The assumption of a constant density of states ensures the
same probability of all final states, which is similar to the equal
{\em a priori} probability in random parking.

The advantage of using the RPP is three-fold. Firstly, the exact
solution for the Fano factor is known analytically
\cite{McKen,Coffman,Bonn}. The numerical value of $F$ in RPP can be
calculated precisely (cf. Eq. \ref{FanoFF}) and is given by \be
F_{exact} = 0.0510387...~. \label{ExactFano}\ee

Secondly, the gap distribution function is also known analytically
(the gaps between cars in RPP are analogous to the kinetic energies
of particles in SEB). This enables us to test the final-state model
hypothesis with exact one-particle distribution function. We
demonstrate that the uncorrelated final-state model gives only a
lower-bound estimate to $F$.

Thirdly, the RPP model has an analytical solution for the
nearest-neighbor two-particle distribution function \cite{Rinto}.
This enables evaluation of the exact correction to the final-state
model due to nearest-neighbor correlations. Inclusion of this
correction gives a close upper-bound estimate of the Fano factor.

\section{Statistical approach }\label{stat}
Statistical evaluation of the Fano factor is based on the analysis
of the full many-particle distribution function in the final state.
Let a particle of initial energy $E$ produce $N$ {\em e-h} pairs of
energies $E_i$  by SEB. The energy balance in the final state is
described by \be E = \sum_{i=1}^N E_i~. \label{balance}\ee It is
convenient to include the bandgap $E_g$ as part of the electron
energy --- both in the final and the initial states; even the
initial energy $E$ is assumed to exceed the kinetic energy by $E_g$
(cf. Appendix A).

The SEB process is terminated  when all $E_i\le E_{th}$, where
$E_{th}$ is the impact ionization threshold energy, i.e. the minimal
energy required to initiate next impact ionization. This is the
final state of SEB and in the RPP model it corresponds to the ``{\em
jamming limit},'' when all remaining gaps are smaller than the car
size.

We assume that $N \gg 1$. This allows us to average  Eq.
(\ref{balance}) over the statistics of SEB. This means the averaging
over a particle set in one realization, which can be taken into
account by replacing $E_i \rightarrow \epsilon$. Next, we average
over multiple realizations of the SEB process, obtaining \be E=
\langle N \rangle \epsilon \label{Av_E}~. \ee

The authors of \cite{Jordan} demonstrated the relation between the
secondary particle energy spread in the final state and the Fano
factor by using an illustrative model called the ``shot-glass''
model. In this model, the SEB process is analogous to filling a
number of small-volume shot-glasses from a bathtub until the latter
is emptied. The individual glass fillings $E_i$ vary randomly with
some distribution, characterized by a mean $\langle E_i \rangle =
\epsilon$ and a variance $\langle \delta \epsilon^2 \rangle =
\langle E_i^2 \rangle - \epsilon^2$.

Consider first the situation when $E$ is not fixed but $N$ is. After
$N$ dippings into the bathtub the amount of water taken from the
tub, $E_N = \sum_1^N E_i$, is a random variable that -- according to
the central limit theorem -- has a Gaussian distribution, \be
P(E_N)=C_N \exp\left[-\frac{(E_N-N \epsilon)^2}{2N \langle \delta
\epsilon^2 \rangle}\right]~, \label{CLT} \ee where $C_N$ is a
normalization constant.

For the case when the total volume ($E$) is fixed, we can
re-interpret Eq. (\ref{CLT}) to give the distribution for the number
of $N$ of filled glasses. Using Eq. (\ref{Av_E}) and neglecting in
the denominator of Eq. (\ref{CLT}) the difference between $N$ and
$\langle N \rangle$, which is a higher-order correction, we find \be
P(N)=C \exp\left[-\frac{\epsilon^2(N-\langle N \rangle)^2}{2\langle
N \rangle \langle \delta \epsilon^2 \rangle} \right]~, \label{CLTN}
\ee where $C$ is another normalization constant. Equation
(\ref{CLTN}) yields the Fano factor in the following form \be
F_{unc} = \frac{\langle  \delta \epsilon^2 \rangle}{\epsilon^2}~.
\label{FanoAv}\ee In the SEB case,  Eq. (\ref {FanoAv}) represents
the Fano factor for uncorrelated particle energy distribution, where
the quantity $\langle \delta \epsilon^2 \rangle$ is the one-particle
energy variance in the final state.

Let us now re-derive an expression for $F$  -- including the
correlation effects. To calculate the deviation of $N$ from its
average for a chosen realization of the SEB process, we rewrite Eq.
(\ref{Av_E}) in the form
 \be  E - N \epsilon = \epsilon(\langle N
\rangle - N) =\sum_{i=1}^N (E_i -\epsilon)~. \label{deviat}\ee Since
the  total energy  is fixed by the initial particle energy, the
spread $\delta N = N - \langle N \rangle$ of the final secondary
particle number results from fluctuations of the secondary particle
energies in the final state. From Eq. (\ref{deviat}) we have \be
(\delta N)^2 \epsilon^2 =\left[ \sum_{i=1}^N
(E_i-\epsilon)\right]^2~, \label{del_N} \ee which is to be averaged
over the statistics in one realization. The result can be written in
the form
\ba \langle (\delta N)^2 \rangle \epsilon^2 =\sum_{i=1}^{N} (\langle
E_i^2 \rangle-\epsilon^2) + \hspace{3cm} \nonumber  \\
 2\sum_{i=1}^{N-1} (\langle E_iE_{i+1} \rangle -\epsilon^2) +
2\sum_{i=1}^{N-2} (\langle E_iE_{i+2} \rangle-\epsilon^2)+...  \nonumber  \\
 + 2\sum_{i=1}^{N-n} (\langle E_iE_{i+n} \rangle-\epsilon^2) +
... ~.  \hspace{3cm} \label{var_energy}  \ea
Equation (\ref{var_energy}) takes into account all possible
correlations between the energies of different electronic pairs. The
right-hand side of (\ref{var_energy}) includes all $N^2$ terms of
the squared sum of the particle energies and is exact.

In the averaging over multiple realizations for large $N$ the sum
$\sum_1^N \langle E_i^2 \rangle$ is self-averaging, viz. \be
\sum_{i=1}^N \langle E_i^2 \rangle = \langle N \rangle \langle E_i^2
\rangle~, \label{S_Av}\ee
and we find
\ba  \langle (\delta N)^2 \rangle \epsilon^2 = \langle N \rangle~( \langle E_i^2 \rangle -
\epsilon^2)  + \hspace{2.5cm} \nonumber \\
+2(\langle N \rangle - 1) (\langle E_iE_{i+1} \rangle - \epsilon^2) + \hspace{2.5cm}\nonumber \\
 +2(\langle N \rangle -2)~(\langle E_iE_{i+2} \rangle - \epsilon^2) + \ldots\hspace{1.9cm}\nonumber \\
+ 2(\langle N\rangle - n)~(\langle E_iE_{i+n} \rangle - \epsilon^2) +\ldots~.\hspace{1.6cm}
\label{var_n} \ea Detailed analysis presented in Sects. 3 and 4 below
shows that the correlations are rapidly decaying with $n$, so that
for large $\langle N \rangle \gg n$ the Fano factor is given by \ba
F = \frac{\langle \delta \epsilon^2 \rangle}{\epsilon^2} +
 \frac{2(\langle E_iE_{i+1} \rangle -\epsilon^2)}{\epsilon^2} +  \ldots \nonumber \\
 + \frac{2 (\langle E_iE_{i+n} \rangle -\epsilon^2)}{\epsilon^2} + \ldots
~. \label{FanoF} \ea
Neglect of all correlations corresponds to retaining only
the first term in the right-hand side of Eq.
(\ref{FanoF}). This reduces (\ref{FanoF}) to Eq. (\ref{FanoAv}).

Note that the use of Eq.(\ref{FanoF}) requires the knowledge of not
only one-particle energy distribution function in the final state,
but also the joint energy distributions for the nearest-neighbor
pairs (corresponding primarily to states created by one impact
ionization), the second neighbors and so on. All of these
distributions essentially define the nature of the final state that
is controlled by the approach to jamming limit.

It is important to emphasize that the above considerations can be
also applied to the intermediate states of the impact ionization
cascade, provided there are enough secondaries for statistics to be
applicable and provided the state evolution is not too fast (the
change in the particle number is smaller than the fluctuations).
This is important because in the real energy branching in $\gamma$
detectors the stationary final state may not be achieved because of
the competing processes of phonon emission (as well as other
particle loss processes, such as recombination and migration to
crystal boundaries). To account for such processes, one must deal
with the intermediate stages of the SEB cascade and, therefore, one
needs to know the time dependence of energy distribution functions.
\section{Kinetic approach }
\subsection{Uncorrelated distribution.}
The RPP model allows an exact evaluation of the distribution of
distances (gaps) between the cars. This can be done by considering
the kinetic (rate) equation that describes the sequential parking
process \cite{RSA,Gonza}. The gap-size distribution function $G(x,t)$
representing the average density of voids of length between $x$ and
$x+dx$ at a time $t$ obeys the following equation \cite{Widom} \be
\frac {\partial G(x,t)}{\partial t} = -k(x)G(x,t)+2
\int_{x+1}^{\infty}dy G(y,t) ~, \label{DistEq} \ee where \be
k(x)=(x-1)\theta(x-1)~, \ee and $\theta(x)$ is the Heaviside step
function. The chosen time scale corresponds to the flux of cars with
1 arrival per unit parking length per unit time.

Equation (\ref{DistEq}) describing the SEB process is a standard
kinetic equation for the energy distribution function of a
homogeneous free electronic gas \cite{Kinetics} where only the
impact collision term is kept. Therefore, Eq. (\ref{DistEq}) can be
easily further specified to include realistic band structure, phonon
scattering, as well as details of the impact ionization process \cite{Boltz}. The
first term in the right-hand side of (\ref{DistEq}) represents
particle loss at energy $x$ due to impact ionization and has a
threshold dependence at $x=1$, the ionization threshold. The second
term corresponds to particle gain due to impact ionization
processes; the factor of 2 reflects the fact that either of the two
secondaries can have the final energy $x$.

\begin{figure}[t]
\epsfig{figure=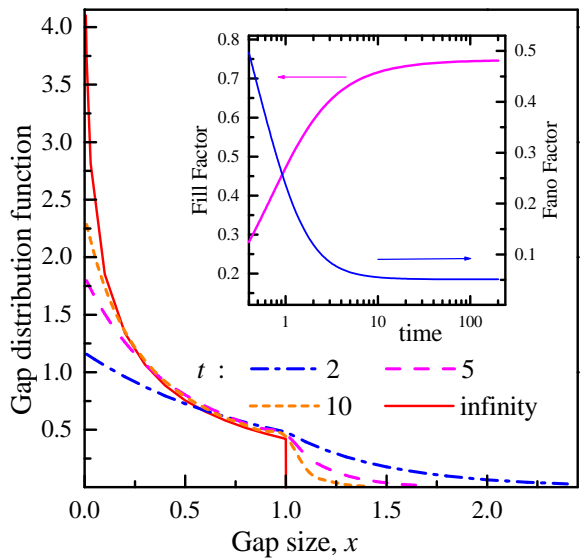,width=7.7cm,height=7.4cm} \caption[]
{(Color online) Evolution with time of the averaged (over realizations) gap
distribution function in the standard random parking problem; the
inset show time variations of both the fill factor (Eq. \ref{rho_t})
and the Fano factor (Eq. \ref{FaTi}). } \label{DFT}
\end{figure}

For an infinite parking lot, Eq. (\ref{DistEq}) can be solved
exactly \cite{Viot} by first seeking the solution at $x>1$ in the
form of a decaying exponent $G(x,t)=f(t) \exp (-xt )$. This yields
\be G(x,t)=t^2\exp\left[-(x-1)t-2\beta(t)\right]~, \hspace{0.5cm}
x>1 \hspace{0.2cm} \ee where \be
\beta(v)=\int_0^vdu\frac{1-e^{-u}}{u}~. \label{beta} \ee Solution
for $x>1$ is then extended to small $x<1$ by using Eq.
(\ref{DistEq}), viz.  \be G(x,t)=2\int_0^t dv v
\exp\left[-xv-2\beta(v)\right]~. \hspace{0.4cm} x<1 \label{Parkdow}
\ee Figure \ref{DFT} shows the evolution of a normalized
distribution $\rho^{-1} G(x,t)$. One observes that the initial
distribution, smooth over a wide range of $x$, evolves into a narrow
distribution within $0<x<1$ interval. The $t \rightarrow \infty$
distribution is dominated by a peak at small $x$ so that the average
gap size is $\approx 0.33$. Temporal evolutions of both the fill
factor and the Fano factor presented in the inset, show very slow
variation from $t=10$ to the jamming state (note the log scale on
the abscissa). Hence the states of main interest are those
immediately preceding the jamming state.

After reaching the jamming limit ($t \rightarrow \infty$), the gap
distribution function becomes
\ba G_\infty(x)=G(x,t)|_{t\rightarrow \infty}=\hspace{3.1cm} \nonumber \\
= 2\int_0^\infty dv v \exp\left[-xv-2\beta(v)\right]~. \hspace{1cm}
x<1~. \label{DistG} \ea Note a logarithmic divergence of
(\ref{DistG}) at small gap values \be G_\infty(x\rightarrow
0)=2e^{-2\gamma}\ln\left(\frac{1}{x}\right), \label{LogDiv} \ee
where $\gamma=0.5772...$ is Euler's constant. The average density of
cars (and of gaps between them) at the time $t$ is given by \be
\rho(t)= \int_0^t dt'  e^{-2\beta(t')}~. \label{rho_t} \ee It can
also be written in terms of a rapidly converging integral,
convenient for numerical evaluation \be \rho(t)= - t e^{-2\beta(t)}
+ 2 \int_0^t dt' e^{-2\beta(t')}e^{-t'} ~. \label{rho_tC} \ee

The growth of $\rho(t)$ saturates at the so-called jamming limit,
when all gaps do not exceed the unit car size. The jamming state
fill factor, \be \rho_\infty = \int_0^\infty dt'
e^{-2\beta(t')}=0.7475987... ~,\label{Fill_f} \ee is known as the
Renyi number.

Next, we use $G(x,t)$ to calculate the uncorrelated Fano factor, Eq.
(\ref{FanoAv}). In terms of the  average gap size $\langle x
\rangle$ the average density of cars $\rho(t) = (1 + \langle x
\rangle)^{-1}$ and \be F_{unc}(t)=\rho(t)^2\langle x^2
\rangle-[1-\rho(t)]^2~. \label{FuPd} \ee Integrating over the gap
distribution in  Eq. (\ref{FuPd}) and using Eq. (\ref{rho_tC}) gives
\ba F_{unc}(t) = 2 \rho(t)[1-\rho(t)] -1 +\hspace{1.5cm} \nonumber
\\*[1ex] 2(1+t)\frac{\rho(t)}{t}e^{-2\beta(t)}+ 4\rho(t)I_{\beta}(t)
\label{FaInt}~, \ea where \be I_\beta(t) = \int_0^t du
e^{-2\beta(u)}\left[\frac {1-(1+u)\exp(-u)}{u^2}\right]
\label{Ibe}~. \ee Numerical evaluation of the integrals in the
jamming state at $t\rightarrow \infty$ gives $F_{unc} \equiv F_{unc}
(\infty) = 0.0439766...$, which is smaller than the exact value, Eq.
(\ref{ExactFano}). The difference is not that large (about 14\%) but
still important so long as the contributions to the Fano factor from
the correlation terms in the right-hand side of Eq. (\ref{FanoF})
are not estimated. In the next section we consider the contribution
of these terms and show that in the jamming limit the
nearest-neighbor correlation corrections are dominant.

Results obtained in this section are strictly valid for parking on
the infinite line. However, we are obviously interested in finite
parking lot lengths, corresponding to SEB of finite initial energy.
The gap distribution function $G_L(x,t)$ suitable for the
formulation of sequential parking  on a line of finite length $L$ is
discussed in Appendix A. In the limit $L\rightarrow \infty$,  due to
the self-averaging property, the function $ G_L(x,t)\rightarrow
G(x,t)$. Numerical experiments show that the two functions are
identical within 1\% already for $L \approx 6$. Therefore, the
results obtained with Eq. (\ref{DistEq}) can be readily used for
finite initial energies.

\subsection{Evaluation of correlation effects}
In the random parking model, evaluation of the correlation
contributions to the Fano factor (Eq. \ref{FanoF}) requires the
knowledge of the pair distribution functions for the
nearest-neighbor gaps, the gaps separated by two cars, three cars
and so on. These are many-particle distribution functions and their
evaluation is not an easy task.

To calculate the first correlation term, one needs the
nearest-neighbor gap distribution function $G_\infty(x,x')$.
Fortunately, this function is known \cite{Rinto}. It can be obtained
as the long-time limit of the time-dependent function $G(x,x',t)$
for which the kinetic equation is of the form
\ba \label{CorrFunc} \frac {\partial G(x,x',t)}{\partial t} = -[k(x)+k(x')]
G(x,x',t) \hspace{1.cm} \nonumber \\
+  \int_{x+1}^{\infty}dy G(y,x',t)+
\int_{x'+1}^{\infty}dy' G(x,y',t) \hspace{1.cm} \\
 + ~ G(x+x'+1,t)~. \nonumber  \hspace{2.5cm} \ea
The source of correlation in Eq. (\ref{CorrFunc}) is seen to be
contained in the last term on the right-hand side, which describes
the appearance of two gaps $x$ and $x'$ upon parking of a car in a
gap of initial length $x+x'+1$. We use the solution  of Eq.
(\ref{CorrFunc}) obtained in \cite{Rinto} to write down the gap pair
distribution function in the final state, \be G_\infty (x,x') =
\lim|_{t \rightarrow \infty}G(x,x',t)~, \label{gapcord} \ee which is
the nearest-neighbor distribution function in the jamming limit,
\be \label{Gxx'}
  G_\infty(x,x')=\int_0^{\infty}dt ~ t^2 e^{-2\beta(t)}e^{-(x +x')t}\ee
$$+ \frac{1}{2} \int_0^{\infty}dt_1  e^{-\beta(t_1)}e^{-xt_1}  \int_0^{t_1}dt_2  e^{-\beta(t_2)}e^{-x't_2} J(t_2) $$
 $$+ \frac{1}{2} \int_0^{\infty}dt_1  e^{-\beta(t_1)}e^{-x't_1}  \int_0^{t_1}dt_2  e^{-\beta(t_2)}e^{-xt_2} J(t_2)~,  $$
where
\be
J(t)=1-e^{-2t}+2te^{-t}~. \label{Jt}
\ee
Similarly to $G_\infty(x)$ in Eq. (\ref{DistG}), the distribution
function $G_\infty(x,x')$ in Eq. (\ref{Gxx'}) gives the number of pairs {\em per unit
length} -- but not the pair probability -- and it must be properly
normalized. By the definition of $G_\infty(x,x')$, the integration
over $x$ and $x'$ gives \ba G_\infty(x) =\int_0^\infty dx'
G_\infty(x,x'), \hspace{2.2cm} \nonumber \\ \rho_\infty =\int_0^\infty
dx\int_0^\infty dx' G_\infty(x,x')~. \hspace{2.1cm} \label{govsgg} \ea Hence, the
normalizing factor is $(\rho_\infty)^{-1}$.

The average two-gap product calculated with the distribution
function $G_\infty(x,x')$ can be written in the form \be
\int_0^\infty dx \int_0^\infty dx'xx' G_\infty(x,x') = K_1+K_2
+K_3~, \label{k1k2k3} \ee
where
\be K_1=\int_0^{\infty}dt ~ t^2e^{-2\beta(t)} I(t)^2~, \label{k1} \ee
\be K_2= \frac{1}{2}\left(\int_0^{\infty}dt_1e^{-\beta(t_1)} I(t_1)\right)^2 ~.
\label{k2} \ee
and
\ba K_3= \int_0^{\infty}dt_1  e^{-\beta(t_1)}I(t_1) \times \hspace{2cm} \nonumber\\
\int_0^{t_1}dt_2 e^{-\beta(t_2)}I(t_2)e^{-t_2}(2 t_2-e^{-t_2}).
\label{k3} \ea Here \be
I(t)=\int_0^{1}dx~x~e^{-tx}\equiv -\frac {d}{dt}\left(\frac
{1-e^{-t}}{t}\right) \label{It}. \ee
Note that both $K_1$ and $K_2$
are positive quantities.  Numerical evaluations of the integrals
gives: \ba \label{Num_K} K_1=& 0.027982,  \nonumber \\
K_2=& 0.072887,  \\ K_3=&-0.010512, \nonumber\ea whence we find that
the additional contribution to $F$ due to nearest-neighbor pair
correlation is given by \ba \label{Num_corF} \delta
F_{nnp}=2\left[\rho_\infty(K_1+K_2+K_3)-(1-\rho_\infty)^2\right] \nonumber\\
= 0.007685  ~. \ea We see that the corrected value of the Fano
factor including nearest-neighbor correlations only,
$F_{nnp}=F_{unc}+\delta F_{nnp} = 0.05208$, is above the exact value
by only 0.001. One can anticipate that in a large parking lot gaps
separated by two or more cars should be only slightly correlated.
Indeed, due to the random nature of parking, only two
nearest-neighbor gaps can be created in a single branching event,
while gaps separated by two cars are created in two random events,
which suggests that their sizes are not correlated. If that were the
case for RPP, then the expansion (\ref{FanoF}) could be restricted
to the nearest-neighbor correlation correction only, so that the
approximation $F=F_{nnp}$ would be exact.

Temporal variation of the Fano factor including nearest-neighbor
correlations only can be found similarly --- with the help of
$G(x,x',t)$ --- but the calculations become rather tedious. As an
example, the values of $\delta F_{nnp}(t)$  for $t=8$ and $t=100$
are, respectively, $\delta F_{nnp}(8)=0.0052649004$ and $\delta
F_{nnp}(100)=0.0074316483$. The calculated results are presented and
discussed below, see Fig. \ref{Compare}.

In fact, all additional terms due to correlations in the positions
of the 2nd, 3rd, ..., neighbors are small but still non-vanishing,
since every division of the parking length imposes restrictions on
the further gap distribution. The next correction $\delta F_{2}$ due
to the 2nd neighbors only is given by an equation similar to
Eqs.  (\ref{k1k2k3}, \ref{Num_corF}), viz.
\ba \label{sec} \delta F_2=2\rho_\infty
\int_0^1\int_0^1dxdx' xx' G_2 (x,x') \nonumber \\\hspace{2cm} -2(1-\rho_\infty)^2~.
 \ea Equation (\ref{sec}) is exact but the pair
distribution function $G_2 (x,x')$ is exceedingly difficult to
calculate because one needs to average an exact three-particle
distribution function over the third particle position. Similar
calculation for more distant pairs would require exact
multi-particle joint distribution functions and averaging over all
intermediate particle positions.

One can still make some progress by taking the {\em factorization
Ansatz} for the multi-particle distributions, expressing them in
terms of nearest-neighbor pair distributions. For the 2nd neighbor
pair distribution function,  this results in the following
approximation
\be G_{2}(x,x')=\int_0^1
G_\infty(x,x'')\frac{G_\infty(x'',x')}{G_\infty(x'')} dx''~,
\label{condi} \ee
where $G_\infty(x'')^{-1}G_\infty(x'',x')\equiv G_c(x'',x')$ is the
conditional probability of finding the second gap equal to $x'$ for
the case when the intermediate gap equals $x''$.

With the help of Eqs. (\ref{condi}, \ref{govsgg}), it is convenient
to rewrite Eq. (\ref{sec}) in the form
\be  \label{convsec} \delta
F_2= 2(1-\rho_\infty) \int_0^1\int_0^1dxdx' x'G_\infty(x,x')
[r(x)-1], \ee
where $r(x)$ is defined in terms of
the ratio of the distributions
\be r(x)=\rho_\infty \frac{ \int_0^1
dx'x' G_c(x,x')}{\int_0^1 dx'x'G_\infty(x')}~. \label{R}\ee
Function $r(x)$ reflects the conditional probability
averaged with weight $x'$ and it approaches unity
when correlations are negligible (for $r=1$ all correlation
corrections vanish).

\begin{figure}[t]
\epsfig{figure=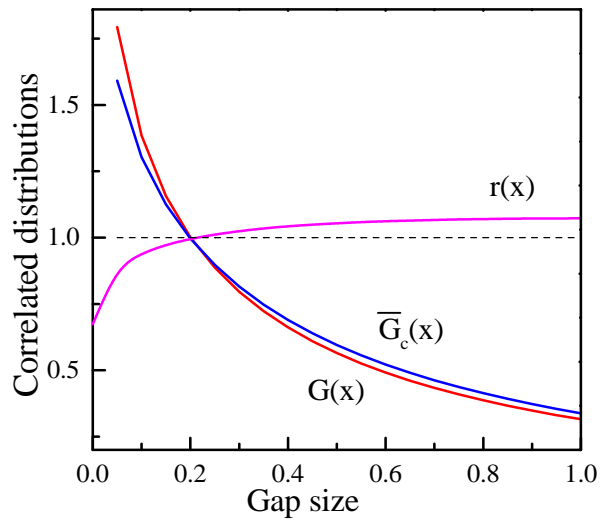,width=7.9cm,height=6.9cm} \caption[]
{(Color online) Distribution ratio $r(x)$ and the conditional probability functions
calculated under the factorization Ansatz (\ref{condi}). }
\label{ratio}
\end{figure}

With the factorization Ansatz, the correlation corrections for more
distant pairs can be similarly expressed through integrals of $r(x)$
over conditional probabilities. Let us see how far we can get with
this Ansatz.

The 2nd neighbor pair correlations described by Eqs. (\ref{condi},
\ref{convsec}, \ref{R}) are illustrated in Fig. \ref{ratio}, which
shows the function $r(x)$ and also compares the functions
$G_\infty(x)$ and \be \overline G_c(x)=
 \rho_\infty \frac{\int_0^1
dx'x'G_\infty(x,x')}{\int_0^1 dx'x'G_\infty(x')}~.
\ee
needed for direct calculation of the first term in the right-hand side of
 Eq. (\ref{convsec}).
We see that both $G_\infty(x)$ and $\overline G_c(x)$ have
a logarithmic singularity at $x\rightarrow 0$. Ratio $r(x)$ is
on average close  to unity and deviates from unity most
noticeably at small $x$ where it approaches the value 0.504.
Numerical calculations give for $\delta F_2=0.0011$ indicating
that the Ansatz series converges.
However, it does not converge to the exact value of $F$. Indeed, the
positive sign of $\delta F_2$ excludes the possibility of reaching
the exact value based on an accurate inclusion of only
nearest-neighbor pair correlations. Evidently, rare multi-particle
correlated configurations become important at this level of
accuracy. The nature of these configurations is discussed in the
next Section.

\section{Discussion}
Analytical expressions for the Fano factor in the jamming state of
the RPP model have been obtained by several authors
\cite{McKen,Coffman,Bonn}. Since these authors used different
techniques (a lattice model was used in \cite{McKen}, a recursive
approach was used in \cite{Coffman}, and a kinetic approach was used
in \cite{Bonn}, whereby $F$ was obtained as a zero-wave-vector
component of the structure factor), their final results were written
in widely different forms, so much so that the equivalence of these
results could be open to question. As shown in Appendix B, the
results of \cite{McKen,Coffman,Bonn} are indeed equivalent and can
be cast in the following rather compact form: \be
F=2\rho_\infty-1-\frac{2}{\rho_\infty}\int_0^\infty
\tilde\rho^2(t)e^{2\beta(t)}A(t)dt ~, \label{FanoFF} \ee where
$\tilde\rho(t)=\rho_\infty -\rho(t)$ and \be \label{At}
A(t)=e^{-t}\left(\frac{e^{-t}+t-1}{t^2}\right). \ee Equation
(\ref{FanoFF}) yields the numerical value of the Fano factor, $F
=0.0510387...$,  that can be calculated with any required precision.

\begin{figure}[t]
\epsfig{figure=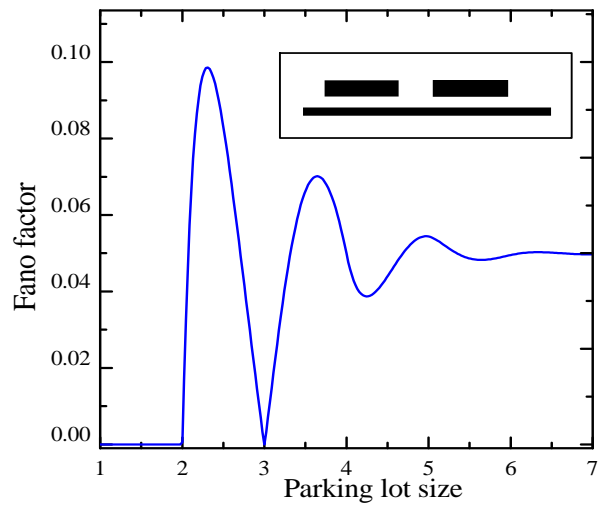,width=7.9cm,height=6.7cm} \caption[]
{(Color online) The Fano factor for the random parking model
calculated as a function of the parking lot length for small lots.
Inset illustrates parking on a lot of length $L = 3$. Even though
the spacing between cars is chosen randomly, there are no
fluctuations of the number of cars. } \label{Fano}
\end{figure}

Numerical evaluation of the Fano factor with account of only the
nearest-neighbor correlation somewhat differs (by 0.0021 or about
4\%) from the exact value. This indicates that, contrary to the
first intuition, distant gap correlations also give a contribution
to $F$. This conclusion is supported by a more refined analysis of
the correlations.  The situation can be clarified by calculating the
variance of $N$ recursively in parking lots of progressively
increasing length. The procedure is described in \cite{Rusb,ASSL}
and here we present (Fig. \ref{Fano}) only the results of
calculations of the Fano factor as a function of the parking lot
length $L$ (avoiding spurious edge effects, as described in Appendix
A). One can clearly see very large variations of the Fano factor for
short parking lot lengths, in the range of up to 5 cars. Such small
gaps appear at an intermediate stage of the parking. These special
correlations are completely smeared out only at $L>5$.

Consider a special case of random parking on a lot whose length is
triple the size of a single car, as shown in the inset of Fig.
\ref{Fano}. One can readily see that two cars will always park in
this lot, with no fluctuation of this number (the unique case of
three tightly parked cars with no gaps has zero probability).
Clearly, the intermediate states of this type were not included in
the preceding consideration and their contribution should reduce the
resulting value of $F$ bringing it to the exact value.

Note that in the jamming state (where the fill factor is close to
3/4) three cars occupy an average length of $L\approx 4$. The
typical space left for 2 cars equals 3 and the above configuration
appears quite common. However, due to the nature of random parking,
these configurations have different pre-histories and most of them
result {\em not} from divisions of $L\approx 3$ lots. The overall
contribution to $F$ of two-car lots remains positive. The negative
contribution results mainly from the those configurations that have
$L\approx 3$ lots in their history. In terms of sequential energy
branching these configurations correspond to an intermediate state
comprising a particle of energy $\approx 3E_{th}$. If such a
particle is created in the course of SEB, the next energy branching
will produce exactly 2 particles with no fluctuation, irrespectively
of the fluctuating kinetic energies of these particles. Inclusion of
this effect is the main residual correction contained in the distant
pair correlation terms. For large initial energies, the overall
contribution of these rare configurations at the jamming state should be
$\delta F_3\approx - 0.0021$.

It would be extremely interesting to realize a situation when
configurations comprising a particle of energy $\approx 3E_{th}$ are
{\em not rare}. For such configurations, the final state will be
dominated by 2-particle contributions. Correlations of this type
will suppress the fluctuations of the final number of particles in
all cases when one of the secondaries produced at an intermediate
stage {\em regularly} has a small energy. One possibility would be
to look for these effects in the dependence of noise in
semiconductor X-ray detectors on the frequency $\nu$ of incident
X-ray flux of constant intensity.  For $h \nu$ producing an initial
electron of energy near $3E_{th}$ one can expect suppression of the
noise component associated with the branching of energy of the
photoabsorbed quanta.

There is also a tantalizing possibility to employ these correlations
in practical  $\gamma$-detectors, where the energy is, of course,
much larger than $3E_{th}$. This possibility relies on the
established fact that in semiconductors the dominant energy loss
mechanism at high electron energies is plasmon emission rather than
impact ionization \cite{Bichsel,Gao}. Plasmon emission can establish
the dominant intermediate configuration --- immediately preceding
the final stage of SEB via impact ionization --- that is populated
with particles of energy close to the plasmon energy, which is
$\approx16$ eV in all common semiconductors. In the RPP language,
the long parking lot, corresponding to the initial energy, would be
divided at the intermediate stage into small 16 eV lots, where ---
as we have seen
--- the small-lot correlations can be very important.

\begin{figure}[t]
\epsfig{figure=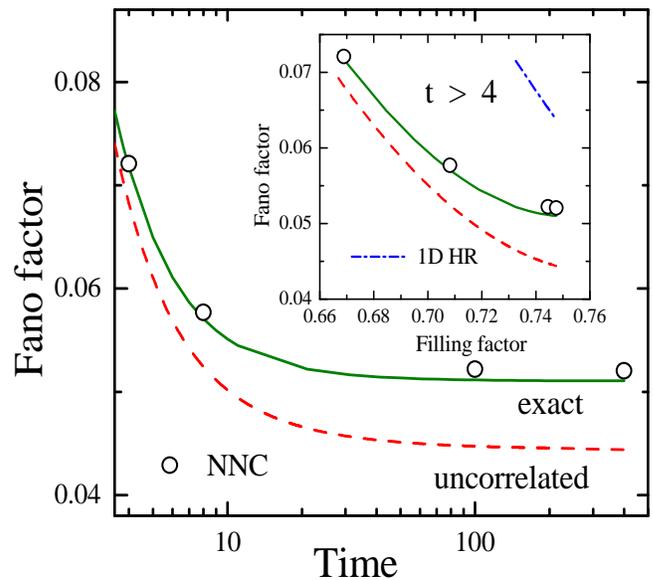,width=8.6cm,height=7.8cm} \caption[]
{(Color online) Fano factor as a function of time for the RPP
problem. The exact result is compared with the shot-glass model
approximation (dashed line) with the exact variance $\langle \delta
\epsilon^2 \rangle$ plugged in. Open circles show the results
obtained by allowing only the nearest neighbor correlations. The
inset show the Fano factor as a function of the line filling factor
in the region approaching the jamming state. It also shows the
predictions of the 1D hard-rod gas model (dash-dotted line in the
top right-hand corner).} \label{Compare}
\end{figure}

As was noted in Sect. \ref{stat}, the kinetic approach allows to
calculate both the filling factor and the Fano factor at any
intermediate state --- by using time-dependent distribution
functions. Figure \ref{Compare} compares the computed values of $F$
for the shot-glass model, where $F_{unc}(t)$ is given by Eq.
(\ref{FaInt}), with exact results (Eq. \ref{FaTi}) for the RPP
model, both as functions of the line filling. The figure also shows
the results obtained including the nearest neighbor correlations.
Allowance for these correlations gives a very close upper estimate
for $F$ for all stages of the random parking.

Inclusion of the correlation corrections obviously requires a
computational effort. We believe it should be quite manageable
because the only important correlations are those preceding the
final state. Nevertheless, the shot-glass model remains attractive
for its simplicity --- as a 0-th order approximation --- even
though its use requires assumptions about
the single-particle distribution that go beyond the model itself. In
this vein, however, there is another statistical model that is,
perhaps, even more attractive.

This model corresponds to a car distribution along the parking lot
in which the probabilities of all allowable states are the same (as
if all cars parked randomly at the same moment). This distribution
is statistically equivalent to the model of a one-dimensional
hard-rod (1D HR) gas, i.e. it can be viewed as an equilibrium
spatial distribution of hard rods of unit length along a segment of
a large total length with a given rod density $\rho$. For the 1D HR
gas model one can calculate all multi-particle distribution
functions exactly \cite{Sals}. It was found that the gaps in the 1D
HR model are distributed in accordance with Poisson statistics and
that $F$ can be exactly expressed in terms of the filling factor
$\rho(t)$, viz. $F = (1-\rho)^2$, see Appendix 3. In the 1D HR
model, there is no jamming limit and the filling factor can take any
value up to $\rho = 1$. The choice of $\rho$, therefore, requires an
assumption that goes beyond the model itself. The 1D HR model gives
a simple way to estimate $F$ --- whenever the filling factor is
known. In a practical application of this model, it is natural to
take the filling factor equal to the branching yield, $\rho =Y$, and
estimate the latter consistently with the average pair excitation
energy $\epsilon$. The 1D HR model gives an upper bound estimate
that is fairly crude compared to the uncorrelated (shot glass)
model, where the exact variance $\langle \delta \epsilon^2 \rangle$
is plugged in.

Both the shot glass and the 1D HR model require assumptions {\em
external} to the model itself: the 1D HR model needs the filling
factor $\rho$, while the shot glass model needs the variance
$\langle \delta \epsilon^2 \rangle$ (derived here from the exact
single-particle distribution).

At small values of $\rho$ all models give $F\approx1-2\rho(t)$. For
intermediate coverage, $\rho \le 0.7$ and near the jamming state,
the uncorrelated model gives a lower bound, while the 1D HR model an
upper bound to the exact $F(t)$. We had already noted (at the end of
Sect. 2) that in a real particle detector, the true jamming state is
hardly achievable because the SEB slows down and is preempted by
phonon emission and other energy loss mechanisms. Owing to the
termination of the branching process at intermediate values of the
filling time, the uncorrelated approximation becomes substantially
less reliable.

 A subtle conceptual problem with the uncorrelated approximation can
be illustrated by a mock shot-glass model. Imagine that the man with
the shot-glass watches what he is doing and compensates for an
underfilled shot by following it with a shot with more than average
filling, so that two successive glasses together scoop similar
amount of water. Evidently, the fluctuations should be strongly
reduced in this case, compared to predictions of Eq. (\ref{FanoAv}).
Moreover, it is precisely these types of correlations that are
typical for the energy branching process. This is easiest seen in the parking model,
where the division of any initial gap naturally pairs a small gap with a large neighbor gap.
We shall refer to this effect as the ``division correlations''.
One could expect that the uncorrelated model --- which ignores the division correlations
--- would give an overestimate of the Fano
factor... but an inspection of Fig. \ref{Compare} reveals that the
opposite is true and the correlation correction to $F$ is positive.
What is wrong with the above argument?

The answer lies in the fact that division correlations are not the
only and not necessarily dominant correlation corrections. Consider
the structure of the terms (\ref{k1}--\ref{k3}). Their physical
meaning is revealed by the corresponding contributions to
$G_{\infty}(x,x')$ of  Eq. (\ref{Gxx'}). From the exponential
dependences on $x$ and $x'$ one can see that the term $K_1$
[originating from the 1st term in $G_{\infty}(x,x')$] depends on the
sum $x+x'$ and thus is manifestly insensitive to division
correlations. Term $K_2$ results from a combination of the 2nd and
the 3rd terms in Eq. (\ref{Gxx'}) with only unity retained from the
$J$ of Eq. (\ref{Jt}). This combination can be factorized and hence
so can be $K_2=\langle x \rangle_{p}^2$, where the subscript $p$
indicates the ``pair averaging'' as in Eq. (\ref{k2}). The
factorizable term  does not represent division correlations either. 
The effect of division correlations resides apparently in the term $K_3$,
which results from the remaining parts of $G_{\infty}(x,x')$. It is
indeed negative but its value is relatively small. As seen from Eq.
(\ref{Num_K}), the term $K_2$ is dominant. Due to the
nearest-neighbor correlations, the pair averaging gives a larger
mean value than the single-particle averaging, $\langle x
\rangle_{p} > \langle x \rangle$. This is undoubtedly  related to
the shape of the single-particle distribution function that peaks at
low $x$.

\section{Conclusions}

We have studied correlation effects in the fluctuation of the number
of particles produced in semiconductor radiation detectors by impact
ionization cascade that leads to sequential energy branching. Our
analysis is based on an exactly soluble random parking model. First,
we show that, in contrast to the so-called ``final state'' models,
the accurate expression for Fano factor includes additional terms
arising from the correlation between energies of the secondary
particles created in the SEB process.  Final state models, such as
the ``shot-glass'' model, are widely used for estimation of the Fano
factor in semiconductors, but they entirely neglect these
correlations. We have considered the best (using an exact gap
distribution function) predictions of the shot-glass model for the
random parking model. We have found that the uncorrelated model
--- even with the exact distribution function --- is not quite accurate
and gives a lower bound to $F$.

Next, we considered the corrections arising from correlations
between nearest-neighbor gaps, next nearest gaps and so on. Based on
the exactly calculated pair distribution function, we found that
nearest-neighbor pair correlations provide the dominant corrections
and their inclusion brings $F$ very close to the exact value. The
residual difference cannot be accounted for by a factorization
Ansatz that expresses distant-neighbor pair distribution functions
in terms of the nearest-neighbor distributions. Instead, one needs
to take into account genuine multi-particle correlations.

The most important example of the correlated configuration that
cannot be factorized into nearest-neighbor correlations is the
intermediate state with the kinetic energy equal 3$E_{th}$ that will
always branch into two particles, with no fluctuation of that number
--- irrespective of the fluctuating energies of these particles. We
have discussed the possibility that this effect may produce an
additional reduction of the Fano factor in semiconductors where the
dominant energy-loss mechanism at high energies is plasmon emission.

More realistic models of energy branching in semiconductor gamma
detectors comprise additional factors (such as non-random branching
at the intermediate stages, energy-loss mechanisms and finite-width
of the valence band) that make the correlation effects different
from those calculated in the random parking model. However, their
importance can be evaluated by the approach developed in this work.

A relatively crude upper  estimate for the Fano factor can be
obtained in the equilibrium statistical model of a one-dimensional
hard-rod gas. The correlations in the 1D HR gas model are somewhat
different from those of sequential energy branching and the model
produces an upper bound to the exact result, provided the filling
factor $\rho$ is known correctly. In this model, the Fano factor has
a simple close-form expression in terms of $\rho$, but the latter is
not limited by jamming and must be determined by considerations
external to the model.
\section*{Acknowledgments}
This work was supported by the Domestic Nuclear Detection Office
(DNDO) of the Department of Homeland Security, by the Defense Threat
Reduction Agency (DTRA) through its basic research program, and by
the New York State Office of Science, Technology and Academic
Research (NYSTAR) through the Center for Advanced Sensor Technology
(Sensor CAT) at Stony Brook.
\appendix
\section{Kinetics for a finite-size parking lot}
Equation (\ref{DistEq}) for RPP and its extensions have been discussed in a
number of papers  (see \cite{Viot} for the review) but only for the
case of {\em parking on an infinite line}. The reason for this
restriction has been that only on the infinite line the number of
voids equals that of the cars and parking of a new car does not
change this property. In a parking lot of finite length the number
of voids exceeds the number of cars by unity, which seemingly makes
the distribution function of voids not suitable for describing the
current number of cars. However, one can consider an initial finite
parking lot of length $L+1$ with one car fixed at the end. Then one
can easily see that the numbers of cars and voids remain equal. The
initial condition to the Eq. (\ref{DistEq}), corresponding to an
empty lot, in this case takes the form \be
G_L(x,t)=\frac{1}{L+1}\delta (x-L)~, \hspace{1cm} t=0~.
\label{t_zero} \ee One can easily check that Eq. (\ref{t_zero})
satisfies the total length conservation condition, \be \int_0^\infty
dx G_L(x,t)+\int_0^\infty dx G_L(x,t)x = 1~. \label{norma} \ee The
time dependent filling factor $\rho(t)$ can either be calculated as
the average number of cars per unit length, \be
\rho_L(t)=\int_0^\infty dx G_L(x,t) \label{Fill_ft}~, \ee or, using
Eq. (\ref{norma}), it can be expressed through the average size of
the gaps.

In the limit $L\rightarrow \infty$,  due to the self-averaging
property, the function $ G_L(x,t)\rightarrow G(x,t)$.

\section{Comparison of expressions for the Fano factor}
An exact expression for the Fano factor in the RPP model was first
derived by McKenzie \cite{McKen}. In the form due to Coffman et al.
\cite{Coffman}, this result can be written as follows \ba
F_C=\frac{4}{\rho_\infty}\int_0^\infty \tilde
\rho(t) e^{-t}\left(\frac{1- e^{-t}}{t}\right)dt \hspace{1cm} \nonumber \\
-\frac{4}{\rho_\infty}\int_0^\infty
\tilde\rho^2(t) e^{2\beta(t)}e^{-t}A(t)dt
  -1~, \label{var_kC} \ea
where $A(t)$ is given by Eq. (\ref{At}). The formula of Bonnier et
al. \cite{Bonn} in the same notations is given by \ba \label{var_kB}
F_B=2\rho_\infty -1 - \frac{4}{\rho_\infty}\int_0^\infty dt
e^{-2\beta(t)}\int_0^t dt' e^{-2\beta(t')} \nonumber \\ \times
\int_0^{t'} dt''e^{2\beta(t'')} A(t'')~.\hspace{1.cm} \ea To prove
their identity, one can use in Eq.(\ref{var_kB}) the substitution
(cf. Eq. \ref{rho_t}) \be \frac {d\tilde \rho(t)}{dt}=- e^{-2
\beta(t)} \label{D_Alp} \ee and then perform integrations by parts.
This brings the integral of Eq. (\ref{var_kB}) into the form of the
second integral in the right-hand side of Eq. (\ref{FanoFF}).
Similar simplification is achieved in the second term of the
right-hand side of Eq. (\ref{var_kC}) by writing
$\exp(-t)=\exp(-t)-1+1$ and then simplifying the term proportional
to $2\exp[2\beta(t)][\exp(-t)-1]/t=d(\exp[2\beta(t)])/dt$ by
integration by parts. The final result of the algebra is again of
the form of Eq. (\ref{FanoFF}). Therefore, both Eqs. (\ref{var_kC})
and (\ref{var_kB}) give identical results.

Evaluation of the Fano factor in the kinetic approach can be
extended \cite{Bonn} to include the temporal evolution of $F$. The
exact $F(t)$ for the RPP model can be written in the form
(simplified in the same way as Eq. \ref{var_kB}): \ba
F_B(t)=2\rho(t) -1+2e^{-2\beta(t)}- \hspace{2cm} \nonumber \\
 -\frac{2}{\rho(t)}\int_0^t dt_1 \tilde\rho(t,t_1)^2 e^{2\beta(t_1)}A(t_1)~, \label{FaTi} \ea
where $\tilde \rho(t,t_1)=\rho(t)-\rho(t_1)$. The  temporal
evolution of $F_B(t)$ is presented in the inset to Fig. \ref{DFT}.

\section{Fano factor for a one-dimensional gas of hard rods}
In this ideal-gas model, the distances between neighboring particles
are distributed according to the Poisson statistics \cite{Ziman},
i.e. \be G(x)dx=\overline x^{-1}e^{-x/\overline x}~dx ~.
\label{poisson} \ee Moreover, in this model there is no correlation
between the gaps separating different particles \cite{Sals} and the
pairwise gap distribution functions can be factored into products of
single-particle functions (\ref{poisson}). Therefore, only the first
term survives in Eq. (\ref{FanoF}). The distribution (\ref{poisson})
gives $\overline x^2-(\overline x)^2= (\overline x)^2$ and for a
given line filling $\rho$ one has $\overline x =(1-\rho)/\rho$, so
that in the notations of Eq. (\ref{FanoF}) $\langle \delta
\epsilon^2 \rangle =(1-\rho)^2/\rho^2 $ and $\epsilon^2=(1+\overline
x)^2=\rho^{-2}$, resulting in \be F_{hr}=(1-\rho)^2 \label{1hD-HR}~.
\ee This result was previously obtained by a much more complicated
calculation. It involves finding the exact pair distribution
function for the rods \cite{Sals} and then calculating  its  Fourier
transform  (the structure factor), see e.g. \cite{Coll}. The Fano
factor is then given  by  the zero-momentum component of the
structure factor. We were able to avoid these complexities by
employing Eq. (\ref{FanoF}) and using the pair distribution function
for gaps separating different particles rather than for particles
themselves.

An alternative way of deriving the Fano factor in the hard-rod gas
model is to use a general expression for the fluctuation of the
number of particles in a given volume \cite{LL}, valid for any
thermodynamic system in equilibrium: \be \langle(\Delta N)^2 \rangle
=- \frac {TN^2}{V^2} \left ( \frac {\partial V} {\partial P} \right
)_{T,N} ~. \label{landau} \ee  For the hard-rod gas, the equation of
state is known exactly (see, e.g., \cite{Coll}), viz. \be P(V-V_0) =
NT ~, \ee giving \be \left(\frac {\partial V} {\partial
P}\right)_{T,N}=- \frac {(V-V_0)^2}{NT}~. \label{dPdV} \ee
Substituting (\ref{dPdV}) into Eq. (\ref {landau}) and taking
$\rho=V_0/V$, we again recover Eq. (\ref{1hD-HR}).


\end{document}